\documentclass[11pt]{article}
\usepackage{epsfig}
\usepackage{amsmath}
\usepackage[small,hang]{caption}
\usepackage{cite}
\usepackage{bm}
\newcommand{\CLrS}{\mathrm{S}}
\newcommand{\CLrA}{\mathrm{A}}
\newcommand{\alphaS}{\alpha_\mathrm{s}}

\newcommand{\CLnumtextS}[1]{\ifcase#1 zero\or one\or two\or three\or four\or
  five\or six\or seven\or eight\or nine\or ten\or eleven\or twelve
  \else #1\fi}
\numberwithin{equation}{section}
\begin{document}
\begin{titlepage}
\hfill
\hspace*{\fill}
\begin{minipage}[t]{4cm}
  DESY--00--038\\
 hep-ph/0003042
\end{minipage}
\vspace*{2.cm}
\begin{center}
  \begin{LARGE}
    {\bf An Estimate of Higher Twist 
      at Small $x_\mathrm{B}$ and Low $Q^2$\\
Based upon a Saturation Model}\\ 
  \end{LARGE} 
  \renewcommand{\thefootnote}{\fnsymbol{footnote}} 
  \footnotetext[0] 
  {Supported by the TMR Network ``QCD and Deep Structure of Elementary 
    Particles'' and by the KBN grant No. 2P03B 089 13.
 One of us (K.G.-B.) is supported by 
    \textit{Deutsche Forschungsgemeinschaft}.} 
  \renewcommand{\thefootnote}{\arabic{footnote}} 
  \setcounter{footnote}{0} 
  \vspace{2.5cm} 
  \begin{Large} 
    {J.Bartels$^{(a)}$, K.Golec-Biernat$^{(a,b)}$ and K.Peters$^{(a)}$}\\ 
  \end{Large} 
  \vspace{0.3cm} 
  \textit{$^{(a)}$II.\ Institut f\"ur Theoretische Physik, 
    Universit\"at Hamburg,\\ Luruper Chaussee 149, 
    D-22761 Hamburg\footnote{email: 
        bartels@x4u2.desy.de $\quad$ krzysztof.golec-biernat@desy.de $\quad$
krisztian.peters@desy.de}           }  \\ 
 \vspace{0.2cm} 
  \textit{$^{(b)}$ Institute of Nuclear Physics, 
   Radzikowskiego 152,\\ 31-342 Krakow, Poland
    } 
\end{center}
\vspace*{1.5cm}
\begin{abstract}
We investigate the influence of higher twist corrections to deep inelastic 
structure functions in the 
low-$Q^2$ and small-$x$ HERA region. We review the general features of the 
lowest-order QCD diagrams which contribute to twist-4 at small-$x$, in 
particular the sign structure of longitudinal and transverse structure 
functions which offers the possibility of strong cancellations in $F_2$.  
For a numerical analysis we perform a twist analysis of the saturation model
which has been very successful both in describing the structure function and
the DIS diffractive cross section at HERA. As the main conclusion, twist
4 corrections are not small in $F_L$ or $F_T$ but 
in $F_2=F_L + F_T$  they almost cancel. This indicates the limitation
on the use of the DGLAP formalism at small $x$ and $Q^2$.
We point out that $F_L$ analysis  needs
a large twist-4 correction. We also indicate the region of validity
of the twist expansion.
\end{abstract}
\end{titlepage}

\section{Introduction}
A deeper understanding of the transition from perturbative QCD to 
nonperturbative 
Pomeron physics in deep inelastic scattering at low $Q^2$ and small $x$ 
remains one of the central tasks in HERA physics. Approaching the transition 
region from the perturbative side, one expects to see the onset of large
perturbative corrections - in particular those which belong to higher twist
operators in QCD. The twist expansion defines a systematic approach and, 
therefore, provides an attractive framework of investigating the region of 
validity of the leading-twist NLO DGLAP evolution equations. The essentials 
of the theory of higher twist operators and their $Q^2$-evolution have been 
laid down twenty years ago: a choice of a complete operator basis has to be 
made ~\cite{EFP}, and for the evolution ~\cite{BFKL} one needs to compute 
evolution kernels which, for partonic operators in leading order, reduce to 
$2\to2$ kernels. The problems of mixing between different operators of
a given twist has also been addressed in          
~\cite{BFKL}. Explicit calculations have been done mainly for
fermionic operators. In the small-$x$ region at HERA, however, we expect 
gluonic operators to be the most
important ones. Recently, a first attempt has been carried out
to analyze the twist-4 gluonic operators in the double-logarithmic
approximation (DLA) ~\cite{BB,BBS}. In addition to analytic calculations also a first
numerical analysis has been presented. As one of the main results,
it has been pointed out that, due to a complicated sign structure, 
subtle cancellations among different twist-4 corrections are possible.
As to the numerical results, the freedom in choosing initial conditions for 
twist-4 gluonic operators, in combination with our presently very limited 
knowledge of the twist-4 evolution equations, make a systematic
QCD study of higher twist corrections in the low-$Q^2$ and small-$x$ region 
at HERA a rather difficult but challenging task.\\ \\ 
In order to gain a detailed insight into the role of higher twist 
it may be helpful to discuss in some detail the  
simplest low-order QCD-diagrams (rather than using the whole 
$Q^2$-evolution machinery collected in ~\cite{BB}). Particular attention
has to be given to the question of possible cancellations between different
contributions.  
As to the choice of input distributions, it seems advisable 
to make use of more specific, model dependent assumptions
on the input distribution. The most reasonable starting point, in our opinion,
is that model which has been most successful in describing the low-$Q^2$ data
of HERA: the saturation model of ~\cite{GBW} 
which contains only four free parameters. This model not only describes
very well the $\gamma^* p$-cross section in the low $Q^2$ transition region 
where the role of higher twist is of particular importance, but also allows 
to connect, in a quantitative way, the total cross section data
with the DIS inclusive diffractive process \cite{GBS}. A particular benefit
of using this model is the interpretation in terms of QCD diagrams:
comparing with the analysis of the QCD diagrams it is possible to
read off a choice of twist-four initial conditions. Since the model 
(before doing any twist expansion) describes the HERA data, it is also likely
to provide a realistic estimate of twist-4 contribution in the low-$Q^2$ and
small-$x$
 region.\\ \\ 
We begin with reviewing the expressions for the simplest 
QCD diagrams and discussing patterns of possible cancellations.
In the following part we review the saturation model and define the twist
expansion. In the third part we perform a numerical analysis and draw 
our conclusions on the the magnitude of gluonic twist-4 corrections.   
The results of our analysis are in qualitative agreement with the estimates
presented in ~\cite{MRST}, suggesting that twist-4 corrections to 
$F_2=F_T+F_L$ are small down to $Q^2 \sim 1~\mbox{\rm GeV}^2$, $x \sim
10^{-4}$. However, we also 
 find that this smallness is due to an almost
complete cancellation 
 of the twist-4 corrections to $F_T$ and $F_L$: both of
them, individually,
 are large, but have opposite signs and nearly the same
magnitude. 
 We interpret this as a warning against using the twist-2
formalism at too low
 $Q^2$ and small $x$.

\section{QCD Diagrams}

\begin{figure}[t]
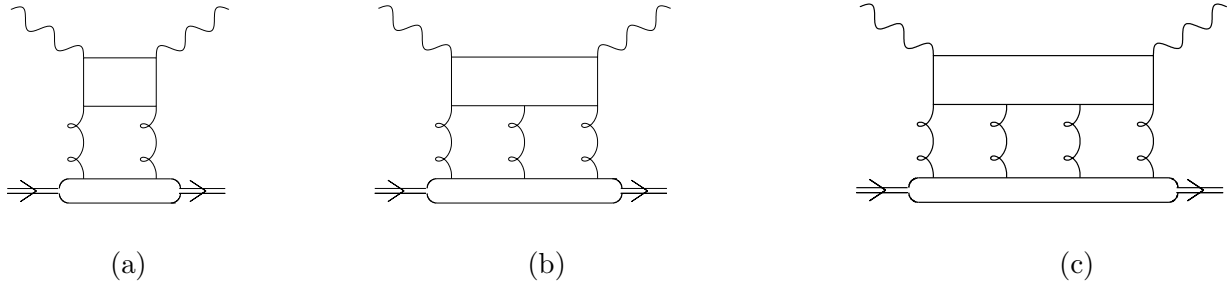
 
\label{figure1}
  \begin{center} 
    \input{2gampl.pstex_t} 
    \input{3gampl.pstex_t} 
    \input{fgleadord1.pstex_t} 
      \end{center} 
        \hspace{1.4cm} (a) \hspace{4.8cm} (b) \hspace{6.3cm} (c) 
      \begin{center} 
    \caption{The simplest QCD diagrams with 2,3 or 4 t-channel gluons.} 
   \end{center} 
\end{figure} 
\begin{figure}[t]
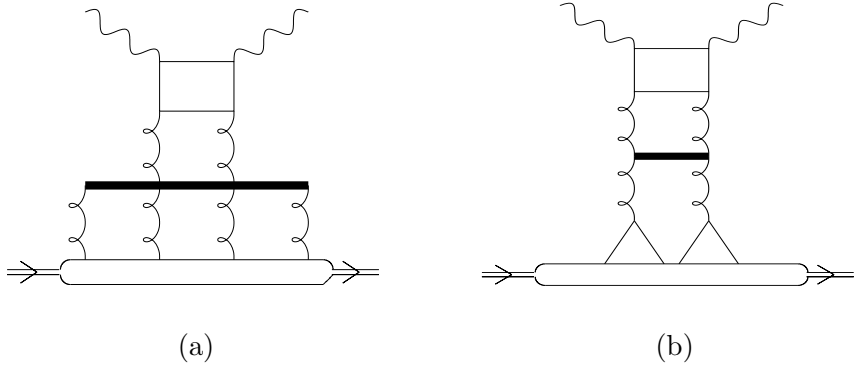

\label{figure2} 
  \begin{center}\hspace{2.5cm} 
    \input{d4ia.pstex_t}\hspace{-1.5cm} 
    \input{d4ra.pstex_t} 
      \end{center} 
         \hspace{4.8cm} (a) \hspace{5.6cm} (b) 
      \begin{center} 
    \caption{Corrections of order $\alpha_s$} 
  \end{center} 
\end{figure} 


We are interested in twist-4 corrections $\Delta F_i$ to the
transverse $(T)$ and longitudinal $(L)$ structure functions
\begin{equation}
F_i(x,Q^2)\;=\;F_i^{\tau=2}(x,Q^2)\;+\;
\Delta F_i(x,Q^2)\,,~~~~~~~~i=T,L.   
\end{equation}
A QCD analysis of twist-4 corrections at small-$x$ starts from the lowest 
order diagrams shown in Fig.~1. The photon can have transverse or 
longitudinal polarization. Approximate expressions for the fermion loop,
$D_{4;0}^{(T,L)(abcd)}$ can be found in ~\cite{BB}: they are valid for small 
$x$, and corrections are of the order $O(x)$. The diagrams with two gluons 
start with leading twist, but they also contain higher twist -  
similar to the BFKL approximation which also contains corrections of all orders
in $1/Q^2$. Unfortunately, we have no way to obtain information on 
the higher-twist couplings to the proton. The diagrams with three gluons, 
through the reggeization of the gluon, are higher order corrections to the 
diagram with two gluons; in the analysis of ~\cite{BBS} they are needed to
complete the covariant derivative of the two gluon diagram. In a complete
twist-4 analysis these diagrams with two and three t-channel gluons have to 
be included, but presently we do not know how to estimate their magnitude. 
The most interesting twist four diagrams, presumably, are the ones with
four gluons. They belong to the four gluon operator which is expected to
play the most crucial role at low $Q^2$ and small $x$. Its coupling
to the proton has been discussed in \cite{BB}, where arguments have been 
given that the simplest model which respects the AGK cutting rules 
consists of (at least) two pieces:
\begin{equation} \label{initialtotal}
  \varphi_4^{abcd}= \varphi_{4\CLrS}^{abcd} + \varphi_{4\CLrA}^{abcd}
\end{equation}
where
\begin{equation} \label{initialsym}
  \varphi_{4\CLrS}^{abcd}=\frac{1}{3\cdot8} \frac{1}{k_1^2k_2^2k_3^2k_4^2}
    \left(
      \delta^{ab}\delta^{cd} f_\CLrS(1,2;3,4;\omega)+
      \delta^{ac}\delta^{bd} f_\CLrS(1,3;2,4;\omega)+
      \delta^{ad}\delta^{bc} f_\CLrS(1,4;2,3;\omega)   \right)
\end{equation}
and
\begin{multline} \label{initialasym}
  \varphi_{4\CLrA}^{abcd}=-\frac{1}{3\cdot8} \frac{1}{k_1^2k_2^2k_3^2k_4^2}
    \Bigl( f^{abm}f^{mcd} f_\CLrA(1,2;3,4;\omega)+ \\
      + f^{acm}f^{mbd} f_\CLrA(1,3;2,4;\omega)+
      f^{adm}f^{mbc} f_\CLrA(1,4;2,3;\omega) \Bigr)\,.
\end{multline}
Here $f_\CLrS$ and $f_\CLrA$ are $\omega$-dependent, positive-valued functions 
which play the role of unintegrated gluon densities. Combining them with the 
quark loop expressions, 
\begin{equation}
\label{patern}
  \Delta F_{T,L}^\mathrm R = -\frac{1}{128\omega \pi^2} 
  \left(\frac{Q^2}{Q_0^2}\right)  
  D_4^{\mathrm R;abcd}\,\otimes \varphi_4^{abcd}(\omega)^{\tau=4}\,,
\end{equation}
and retaining only those terms
which give rise to $Q^2$-logarithms, we arrive at the following twist-4
corrections to the transverse and longitudinal structure functions:
\begin{align} \label{deltaFR}
  \Delta F_T^\mathrm R 
  = \frac{1}{64}\,\frac{\alphaS^2}{\pi^2}
\,
  \sum_f e_f^2\; \frac1\omega  \left( \frac{Q_0^2}{Q^2}
\right)\frac{2}{5}\cdot
     \frac{1}{3} [14\varphi_{4\CLrS}(\omega)-
9\varphi_{4\CLrA}(\omega)] \,.
 \end{align}
and\footnote{We changed the definition of $F_L$ in comparison to 
              \cite{BB}. Now, $F_L$ is twice the previous one to have
              $F_2=F_T+F_L$}
\begin{align} \label{deltaFRl}
  \Delta F_L^\mathrm R = -\,\frac{2}{64}\,\frac{\alphaS^2}{\pi^2}\,
  \sum_f e_f^2\; \frac1\omega  \left(\frac{Q_0^2}{Q^2}\right)
   \left(\frac{94}{225}
    +\frac{4}{15}\ln\left(\frac{Q^2}{Q_0^2}\right)\right)
    \frac{1}{3} [14\varphi_{4\CLrS}(\omega)- 9\varphi_{4\CLrA}(\omega)] \,.
\end{align}
The $\omega$-dependence of initial conditions $\varphi_{4\CLrS}(\omega)$
and $\varphi_{4\CLrA}(\omega)$ will be assumed to lead to a power-like
behavior of the form $(1/x)^{2\lambda}$ where the exponent $\lambda$ is 
unknown. Thus, together with the
$1/Q^2$ suppression, these twist-4 corrections are of the form
\begin{equation}
\label{xdep4}
\Delta F_{T,L} \sim \frac{Q_0^2}{Q^2} \left(\frac{1}{x}\right)^{2\lambda}.
\end{equation}
From this general observation one immediately sees that the value of $Q^2$
where twist four becomes important is $x$-dependent.  
One of the most striking features is the sign structure: the transverse and 
longitudinal cross sections, (\ref{deltaFR}) and (\ref{deltaFRl}), 
have opposite signs, and in $\Delta F_2=
\Delta F_T+\Delta F_L$ 
one faces a strong cancellation. If $\varphi_{4\CLrS}$ and 
$\varphi_{4\CLrA}$ are of the same order of magnitude (such that the
square bracket expression is positive), we expect the twist-4 corrections
to $F_2$ being slightly dominated by the negative corrections to $F_L$, i.e.
the higher twist corrections to $F_2$ could be small and negative.

It is important to note that these corrections to the deep inelastic 
structure functions are closely related to the twist-4 corrections to the 
cross section of diffractive $q\bar{q}$ production 
(the s-discontinuity line between gluon
2 and 3). Twist-4 in diffractive $q\bar{q}$ production has been observed 
experimentally, and it has to be a part of the twist-4 corrections to 
$F_T$ or $F_L$. However, contrary to the most naive expectation, the
two-gluon 
 systems on both sides of this cutting line cannot be restricted to
be in 
 color singlet states: the AGK rules in perturbative QCD ~\cite{BR}
require
 a structure of the form (2.2) and (2.3), and from this one easily
sees that the system 
 of gluons 1 and 2 is not restricted to color singlet.

Turning to corrections of the form $\alpha_s \ln Q^2/Q_0^2$ to Fig.~1  we
face 
 a mixing problem. In \cite{BB} it has been argued, on the basis of an
all-order analysis in the double-logarithmic approximation, that there are
several gluonic twist-4 operators, four-gluon operators and one two
gluon operators. They have different $Q^2$-evolution equations, and they
couple to the proton with different couplings. Furthermore, they mix, 
i.e. there are transitions from two-gluon states in the t-channel to 
four-gluon states. When computing $\alpha_s \ln Q^2/Q_0^2$ corrections to 
the diagrams in Fig.~1(c), we  group the contributions in the form shown in
Fig.~2:  
(a) illustrates the transition of the two-gluon state to the 
four-gluon state and leads to the correction $\Delta F^\mathrm I$, 
and (b) illustrates the first evolution step of the 
 two-gluon operator
 and gives the correction to $\Delta F^\mathrm R$. The
calculation of these diagrams is described in 
 \cite{BB} and leads to the
following twist-4 contributions:
 
\begin{align} \label{D4final} 
  \Delta F_T^\mathrm I = 
    -\frac{1}{16}\, \frac{\alphaS^3}{\pi^3}\, 
    \sum_f e_f^2\; \frac{1}{\omega^2}\left(\frac{Q_0^2}{Q^2}\right) 
    \ln\left(\frac{Q^2}{Q_0^2}\right)\frac{2}{5}  
    \varphi_{4\CLrS}(\omega)\,,  
\end{align}
\begin{align}  
  \Delta F_L^\mathrm I = 
    \frac{2}{16}\, \frac{\alphaS^3}{\pi^3}\, 
    \sum_f e_f^2\;\frac{1}{\omega^2}\left(\frac{Q_0^2}{Q^2}\right) 
    \left\{\frac{94}{225}\ln\left(\frac{Q^2}{Q_0^2}\right)+ 
          \frac{4}{15}\ln^2\left(\frac{Q^2}{Q_0^2}\right)\right\} 
    \varphi_{4\CLrS}(\omega)\,,  
\end{align}  
and
\begin{align} 
\label{deltaFR1}
  \Delta F_T^\mathrm R = \frac{1}{64}\,\frac{\alphaS^2}{\pi^2}\,
  \sum_f e_f^2\; \frac1\omega  \left(\frac{Q_0^2}{Q^2}\right)   
\left(
1+\frac{N_c\alpha_s}{\pi\omega}\ln\left(\frac{Q^2}{Q_0^2}\right)
\right)
\frac{2}{5}\cdot 
\frac{1}{3} [14\varphi_{4\CLrS}(\omega)- 9\varphi_{4\CLrA}(\omega)] \,, 
\end{align}
\begin{align}
\label{deltaFRl1}  
  \Delta F_L^\mathrm R &= - \frac{2}{64}\,\frac{\alphaS^2}{\pi^2}\, 
  \sum_f e_f^2\; \frac1\omega  \left(\frac{Q_0^2}{Q^2}\right) 
\left\{
\frac{94}{225}\left[1+\frac{N_c\alpha_s}{\pi\omega}
\ln\left(\frac{Q^2}{Q_0^2}\right)\right]+ 
\frac{4}{15}\left[\ln\left(\frac{Q^2}{Q_0^2}\right)
+\frac{N_c\alpha_s}{\pi\omega}\ln^2\left(\frac{Q^2}{Q_0^2}\right)\right]
\right\}
\notag \\      
&~~~~~~~~~~~~~~~~~~~~~~~~~~~~~~~~~~~\times \frac{1}{3}
\left[14\varphi_{4\CLrS}(\omega)- 9\varphi_{4\CLrA}(\omega)\right] \,.  
\end{align}  
Note by the comparison of (\ref{deltaFR1}), (\ref{deltaFRl1}) with
(\ref{deltaFR}), (\ref{deltaFRl}),  respectively, that $\Delta F^R$ gets
additional $\alpha_s \ln Q^2/Q_0^2$ corrections 
without changing the structure of initial conditions. The
corrections $\Delta F^I$ are  of the order $\alpha_S^2\,(\alpha_S\ln
Q^2/Q_0^2)$, thus they are not present in the lowest order result which is
proportional to $\alpha_S^2$.  

For low $Q^2$-values it is not a priori clear whether these corrections
to
 the twist-4 contributions are important or not: there is an
additional suppression factor
 $\frac{N_c \alpha_s}{\pi}$, and for low
$Q^2$-values the logarithm $\log Q^2/Q_0^2$ 
does not
 provides much enhancement. To get a first
idea, it may, again, be useful to draw a connection with diffractive 
dissociation.  As illustrated in Fig.~2(a), these diagrams describe diffractive
production of $q\bar{q}g$ systems. There is no doubt that these diffractive 
states have been observed at HERA: a direct analysis of their twist-4 
component (e.g. the observation of diffractive final states with only 
 hard
jets) would provide a direct evidence for the presence of these higher 
 twist
corrections in the deep inelastic structure function.\\ \\
A simple analysis of twist-4 corrections could be based upon the presented
low-order
 expressions. However, even within this framework we need two
initial 
conditions, $\phi_S$ and $\phi_A$. Relating them to the twist-4
diffractive 
 $q\bar{q}$ cross section (as described in \cite{BB}) gives only
one condition,
 and, hence, is not enough. We are therefore lead to build a
model for the 
 initial conditions. The most successful description of the
low-$Q^2$ transition
 region at HERA has been provided by the saturation model 
of \cite{GBW}, and we will use this model to determine the 
initial conditions. 
    
\section{Twist Four in the Saturation Model}

Let us first briefly review the model
of \cite{GBW}
and its decomposition into twist components. It is well known that    
the $\gamma^* p$-cross sections, 
\begin{equation}
\label{sat0}
\sigma_{T,L}(x,Q^{2})\,=\, \frac{4\pi^2\alpha_{em}}{Q^2} F_{T,L}(x,Q^2)\,,
\end{equation} 
can be written at small $x$ as \cite{NZ,FR}:
\begin{equation}
\label{sat1}
 \sigma_{T,L}(x,Q^{2})\,=\,\int d^2\textbf{r} \int_0^1 dz\, |\Psi_{T,L}
(z,\textbf{r})|^{2}\; \hat{\sigma}(x,r^2)
\end{equation}
where $\Psi_{T,L}(z,\textbf{r})$ denotes the transverse  and 
longitudinally polarized photon wave functions, and $\hat{\sigma}(x,r^2)$
is the  dipole cross section which describes the interaction of the
$q\bar{q}$ pair with the proton. In addition,  $z$ is the 
 momentum fraction
of the photon carried by the quark, and $r$ is the relative
 transverse
separation between the quarks. The wave functions are solely 
 determined by
the coupling of the photon to the $q\bar{q}$ pair, see e.g. \cite{FR}. 
 In
\cite{GBW} the dipole cross section
is assumed to depend on $x$ through
the ratio  of the transverse separation $r$ and
the saturation radius $R_0(x)$, and the following form is proposed: 
\begin{equation}
\label{sat2}
\hat{\sigma}(x,r^2)\,=\,\sigma_0\,g\left(\frac{r^2}{4R_0^2}\right)\,\equiv\,
\sigma_0 \left\{1-\exp\left(-\frac{r^2}{4R_0^2}\right)\right\}\,.
\end{equation}
At small $r$ ($r\ll 2R_0$),  
the dipole cross section grows quadratically with 
$r$, $\hat{\sigma}\sim\sigma_0 r^2/4R_0^2$, while for large $r$ ($r \gg 2R_0$),
it saturates, $\hat{\sigma}=\sigma_0$. In order to describe the energy
dependence
 both of the total DIS cross sections and of the low-$Q^2$
cross section 
 measured at HERA, the saturation radius has the following 
$x$-dependent form:
 
\begin{equation}
\label{sat3}
R_0^2(x)\,=\,\frac{1}{Q_0^2} \left(\frac{x}{x_0}\right)^{\lambda}
\end{equation}
with $Q_0 = 1~\mbox{\rm GeV}$. 
Eqs.~(\ref{sat1})-(\ref{sat3}) define the saturation model.
The physical
motivation for such a parameterization and its significance for diffractive
processes in DIS is discussed at length in \cite{GBW,GBS}.
The three parameters in the model are determined from a fit 
to the total DIS cross section data at $x<0.01$ and look as follows:
$\sigma_0 = 23$~\mbox{\rm mb}, $x_0 = 3\cdot 10^{-4}$ and $\lambda = 0.29$. 
In this way a very good
description of data in a broad range of $Q^2$ and $x$ is obtained. 
In fact there is a fourth parameter in the model, 
an effective 
quark mass $m_f=140~\mbox{\rm MeV}$ in the photon wave
function, chosen such that the results of the model, extended down to
photoproduction region, are in a good agreement with photoproduction data 
measured at HERA.

In order to evaluate the cross section (\ref{sat1}) it is convenient to
employ the Mellin 
 transform to factorize the wave function from the
dipole cross section:
 
\begin{eqnarray}
 \nonumber  
\label{mellinint1}
\sigma_{T,L}(x,Q^2)
&=&
\int_{-\infty}^\infty \frac{d\nu}{2\pi} \int d^2 \textbf{r} \int_0^1 dz
|\Psi_{T,L}(z,\textbf{r})|^2 \int \frac{dr^{\prime2}}{r^{\prime2}} 
\left(\frac{r}{r^{\prime}}
\right)^{1+2i\nu} \sigma(x,r^{\prime2}) 
\\ \nonumber  
\\
&=& 
\sigma_0 \int_{-\infty}^\infty \frac{d\nu}{2\pi} 
\left( \frac{1}{Q^2 R_0^2(x)} \right)^{1/2+i\nu} 
H_{T,L} \left( \nu, \frac{m_f^2}{Q^2} \right) G (\nu)\,.
\end{eqnarray}  
In the case of zero quark mass we obtain
\begin{equation}
H_T(\nu,0)=\frac{6\alpha_{em}}{2\pi}\, \sum_f e_f^2\, 
\frac{\pi}{16} 
 \frac{9/4+\nu^2}{1+\nu^2} \left( \frac{\pi}{\cosh(\pi\nu)}
\right)^2 
 \frac{\sinh(\pi\nu)}{\pi\nu}
\frac{\Gamma(3/2+i\nu)}{-\Gamma(-1/2-i\nu)}
 \end{equation}
and
\begin{equation}
H_L(\nu,0)= \frac{6\alpha_{em}}{2\pi}\,  
\sum_f e_f^2\,\frac{\pi}{8}
\frac{1/4+\nu^2}{1+\nu^2} \left( \frac{\pi}{\cosh(\pi\nu)} \right)^2 
\frac{\sinh(\pi\nu)}{\pi\nu} \frac{\Gamma(3/2+i\nu)}{-\Gamma(-1/2-i\nu)}.
\end{equation}
In addition, $G(\nu)$ in (\ref{mellinint1}) equals:
\begin{equation}
G(\nu)= \int_o^\infty d\hat{r}^2 \left( \hat{r}^2 \right)^{-3/2-i\nu} g(\hat{r}
^2) = -\Gamma(-1/2-i\nu),
\end{equation}
for the saturation form of the dipole cross section 
$g(\hat{r}^2)=1-e^{-\hat{r}^2}$. 
From relation (\ref{mellinint1}) we see that the cross sections
$\sigma_{T,L}$ depend on $x$ in the saturation model only 
through the combination 
\begin{equation}
\label{mellinint2}
\xi\,\equiv\,\frac{1}{Q^2 R_0^2(x)}
\,=\,\frac{Q_0^2}{Q^2}\left(\frac{x_0}{x}\right)^{\lambda}.
\end{equation}
Notice that a similar combination, eq.~(\ref{xdep4}), occurs in the twist-4
analysis in the previous section.
\begin{figure}
\label{figure3}
  \begin{center} 
     \input{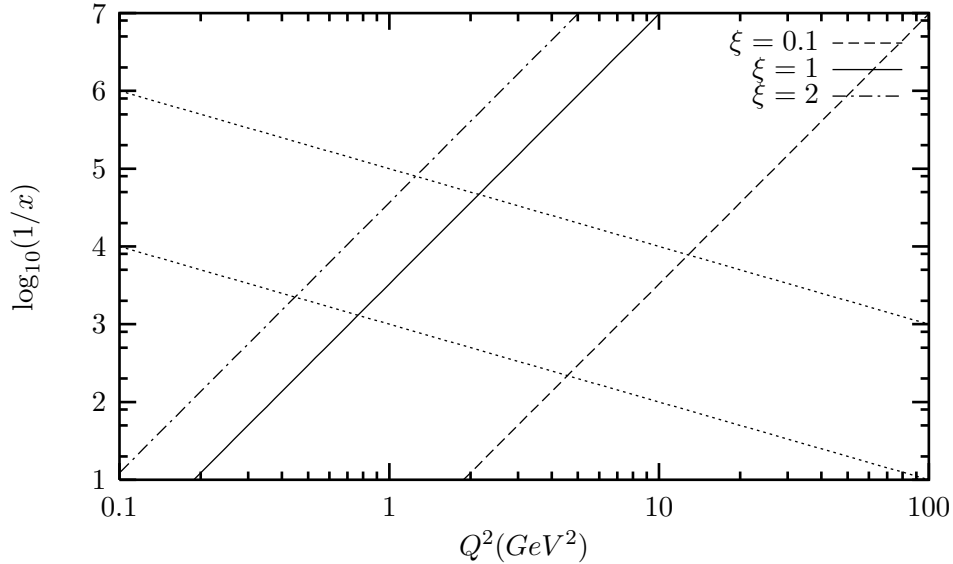} 
       \caption{The solid line ($\xi=1$) in the $(x,Q^2)$ plane indicates a
                critical line of the model \cite{GBW}. 
                The area between the two
                dotted lines corresponds to the acceptance region of HERA.}
 
       \label{10}
\end{center}
\end{figure}

The detailed discussion in ~\cite{GBW} (which will not be repeated here) 
shows that there is an essential change in the energy (and also
$Q^2$) dependence
 of the cross section if we move from the ``perturbative''
region, $\xi<1$
 (large $Q^2$ and not too small $x$), to the nonperturbative
``Pomeron'' 
 region, $\xi>1$ (small $Q^2$ and very small $x$). In the former
region we have
 a power-like rise in $1/x$ (in agreement with the observed
rise of the 
 structure function at small $x$), whereas in the latter one the
cross section 
 stays constant (apart from a logarithmic enhancement factor).
Hence the region 
 $\xi \approx 1$  (see Fig.~3) marks the transition from one
region to the other. 
 In the region $\xi<1$ it is natural to expand the
inclusive cross sections in powers of $\xi$ 
(which means powers of $1/Q^2$)
while in the region $\xi>1$ in powers of 
$1/\xi$ or $Q^2$. It is clear, however, that the expansion
in powers of $1/Q^2$ cannot be valid down to $Q^2=0$. Therefore,
approaching the transition region $\xi \approx 1$
from the perturbative side one expects the 
leading-$\xi$ approximation to fail somewhat to the right of the line 
$\xi=1$ (called a critical line in \cite{GBW}).

With expressions (\ref{mellinint1})-(\ref{mellinint2})  the  above discussion
can be made more precise. 
 The expansion in powers of $\xi$ will be
identified as ``twist expansion''
 (below we will explain why). The power
series in $\xi$ (or $1/\xi$) is 
 determined by the singularities of the
integrand in eq.~(\ref{mellinint1}) in the complex $\nu$-plane. These are
single or multiple poles, located at $\nu=\pm i (2n+1)/2$ with $n=0,1,2,...$.

The $\nu$-integration runs along the real axis, and
for $\xi <1$  it is tempting 
to close the contour in the lower half-plane. A closer look reveals that
in such a case 
an asymptotic expansion for our cross sections is constructed. 
This is done by computing residues of the poles in the lower half-plane
which leads to an expansion in powers of $\xi$ or $1/Q^2$.
Now, 
the critical line (or better, a strip) indicates a limit on validity
of the asymptotic twist expansion. It is  also possible to close the contour
in the upper half-plane in which case an convergent
expansion in positive powers of
$1/\xi$ or $Q^2$ is obtained for
any value of $\xi \ne 0,~\infty$\footnote{We thank Lech Mankiewicz for
a discussion on this point.}. However, this expansion is not practical
in the large $Q^2$ or small $\xi$ analysis.

The first singularity encountered in the lower plane 
is a pole at $\nu=-i/2$. In the saturation
model the transverse cross section has a double pole which generates a 
logarithmic behavior for the leading-twist contribution:
\begin{equation}
\label{twist2t}
\sigma_T=\sigma_0 \sum_f e_f^2 \frac{\alpha_{em}}{\pi} \left( \frac{7}{6} \xi
- \psi(2)\xi+\xi\ln(1/\xi) \right),
\end{equation}
The longitudinal leading-twist contribution has only a 
single pole and therefore does not produce a logarithm: 
\begin{equation}
\label{twist2l}
\sigma_L=\sigma_0 \sum_f e_f^2 \frac{\alpha_{em}}{\pi} \xi.
\end{equation}

Higher-twist contributions can be obtained by evaluating the residues 
at the lower lying poles , the
twist-4 contribution 
 at the pole $\nu=-3i/2$, etc.  
The results for the transverse contributions are:\\ \\
Twist-4:
\begin{equation}
\label{twist4t}
\sigma_T=\sigma_0 \sum_f e_f^2 \frac{\alpha_{em}}{\pi}
\frac{6}{10}\xi^2,            
\end{equation}
Twist-6:
\begin{equation}
\label{twist6t}
\sigma_T=\sigma_0 \sum_f e_f^2 \frac{\alpha_{em}}{\pi}
\left( \frac{43}{1225} \xi^3 - \frac{12}{35} \psi (4) \xi^3 +\frac{12}{35}
\xi^3 \ln(1/\xi) \right),
\end{equation}
Twist-8:
\begin{equation}
\label{twist8t}
\sigma_T=\sigma_0 \sum_f e_f^2 \frac{\alpha_{em}}{\pi}
\left(- \frac{262}{11025} \xi^4 + \frac{4}{35} \psi (5) \xi^4 -\frac{4}{35}
\xi^4 \ln(1/\xi) \right).
\end{equation}
For the longitudinal contributions we find:\\ \\
Twist-4:
\begin{equation}
\label{twist4l}
\sigma_L=\sigma_0 \sum_f e_f^2 \frac{\alpha_{em}}{\pi}
\left(- \frac{94}{75} \xi^2 + \frac{4}{5} \psi (3) \xi^2 -\frac{4}{5}
\xi^2 \ln(1/\xi) \right),
\end{equation}
Twist-6:
\begin{equation}
\label{twist6l}
\sigma_L=\sigma_0 \sum_f e_f^2 \frac{\alpha_{em}}{\pi}
\left( \frac{654}{1225} \xi^3 - \frac{36}{35} \psi (4) \xi^3 +\frac{36}{35}
\xi^3 \ln(1/\xi) \right),
\end{equation}
Twist-8:
\begin{equation}
\label{twist8l}
\sigma_L=\sigma_0 \sum_f e_f^2 \frac{\alpha_{em}}{\pi}
\left(- \frac{1636}{18375} \xi^4 + \frac{48}{175} \psi (5) \xi^4 -\frac{48}
{175}\xi^4 \ln(1/\xi) \right).
\end{equation}

It is not surprising to see a strong similarity between the
twist-4 contributions  (\ref{deltaFR}) and  (\ref{deltaFRl}) computed to
lowest order in QCD, and those found in the saturation model 
(\ref{twist4t}) and  (\ref{twist4l}). In particular, the sign structure is
the same: $\sigma_T$ is positive while $\sigma_L$ is negative. Also,
$\sigma_L$ contains a logarithm while $\sigma_T$ does not. Finally,
the ratio of the twist-4 contributions,
\begin{equation}
\label{twist4ratio}
\frac{\sigma_L}{\sigma_T}\,=\,-\frac{{94}/{75}\,+\,{4}/{5}\,( \ln(1/\xi)\,-\,
\psi(3))}{{3}/{5}}\,,
\end{equation}
is similar to the  ratio obtained from the lowest order
QCD calculation
\begin{equation}
\frac{\Delta F^R_L}{\Delta F^R_T}\,=\,
-\frac{{94}/{75}\,+\,{4}/{5} \ln(Q^2/Q_0^2)}{{3}/{5}}\,.
\end{equation}

Thus, the saturation model can be
interpreted as a result of summing the diagrams shown in Fig.~1(a)
 and 1(c)
(and more ``iterations of gluon ladders''): the leading twist  
 comes
entirely from Fig.~1(a), twist four from Fig.~1(c) etc. 
Viewed in this way,
 the model can be
used to define values of the initial conditions for (\ref{deltaFR}) and 
(\ref{deltaFRl}),  but only in 
 the
combination $14 \varphi_{4\CLrS} - 9\varphi_{4\CLrA}$.
 The higher order
corrections illustrated in Fig.~2(a) are not included in the model. The
success 
 of the model might indicate that this type of corrections is not
very 
 important near the transition (critical) line; but it may also be that a
suitable 
 modification of the model might be necessary.

\section{Numerical Analysis}

Further insight will be gained through a numerical study. We choose to use 
the expressions of section 3 and compare the full expressions
(\ref{mellinint1}) (with $m_f=0$), which can be integrated numerically, 
with the twist-expansion (\ref{twist2t})-(\ref{twist8l}).
As far as the twist-4 terms alone are concerned, we could have 
started from the saturation model, extract the initial conditions and than 
turned to the QCD approximations listed in section 2. For our present 
discussion, however, we find it more instructive to study the role of 
higher twist (in particular, twist-4) in a more general context.
The saturation model, which describes the data, provides an analytical formula
for the cross sections and allows to investigate the twist expansion and its
breakdown near $\xi=1$.
Since the dependence upon $Q^2$ and $x$ is through 
the variable $\xi$, we present the numerical results as a function of $\xi$.

In Fig.~\ref{figure3} we show lines of constant $\xi$ ($\xi=2,\;1,\;0.1$):
variation of 
 $\xi$ means moving from one line to another. 
 With the help of Fig.~3 $\xi$ can be translated into 
the $x$ and $Q^2$ variables.
We begin with presenting the ratios of 
 the leading-twist approximations to the full cross
sections as a function of $\xi$. 
In Fig.~4
 we show the ratios for the transverse
$\sigma_T^{\tau=2}/\sigma_T$,
longitudinal $\sigma_L^{\tau=2}/\sigma_L$,
and 
the total 
$(\sigma_L^{\tau=2}+\sigma_T^{\tau=2})/(\sigma_T+\sigma_L)$, cross sections.
The striking result is that for both the transverse and longitudinal
cross 
 sections separately the higher twist corrections become large when we
are approaching  the transition line $\xi=1$. Moreover, the 
transverse higher twist correction is positive while the
longitudinal one is negative. In the sum, however,
those corrections almost cancel each other and the overall correction 
is small. 

\begin{table}[t]                          
\begin{center}                            
  \begin{tabular}{|l|l|}\hline            
           $x$ & $D_2(x) (GeV^2)$\\ \hline
             0-0.0005 & 0.0147 \\    
             0.0005-0.005 & 0.0217 \\
             0.005-0.01 & -0.0299 \\       
             0.01-0.06 & -0.0382 \\        
             0.06-0.1 & -0.0335 \\         
             0.1-0.2 & -0.121 \\           
             0.2-0.3 & -0.190 \\           
             0.3-0.4 & -0.242 \\           
             0.4-0.5 & -0.141 \\ \hline  
 \end{tabular}                                                           
    \caption{The MRST values of the higher-twist coefficient $D_2(x)$
      in Eq.~(\ref{mrstf2})}
\end{center}
\end{table}

This is a possible explanation of the smallness  of the higher twist
correction to $F_2$ found in the analysis of MRST \cite{MRST}. 
The authors of this analysis use the following simple parameterization: 
\begin{equation} 
\label{mrstf2}
F_2^{HT}(x,Q^2)=F_2^{LT}(x,Q^2)\left(1+\frac{D_2(x)}{Q^2}\right)\,, 
\end{equation} 
and determined the function $D_2(x)$ from a fit to DIS data. 
The result is given in Table~1 which we reproduce  from \cite{MRST}.
 In general, for small $x$,
$D_2(x)$ is small and negative but it becomes positive for the smallest
$x$. We found a similar result, the leading twist approximation  
deviates from the exact formula by less than 10\%. 
The sign structure of this deviation also agrees with  the MRST analysis; 
it is negative but to the left
of the transition line the deviation becomes
positive. The last result should be taken with some care  since, strictly
speaking, the twist expansion for $\xi>1$ in the saturation model makes no
longer sense,  and a new expansion in powers $Q^2$ is appropriate. We
extrapolated, however, the leading twist formula to that region and found an
agreement with the exact result up to $\xi=2$. 
This indicates that the phenomenological success of a leading-twist
analysis might be deceptive: the leading-twist approximation to $F_2$  
remains a good approximation also in the region in which the whole twist
expansion has already  collapsed.

The last point is also illustrated in Fig.~5 
where  the individual higher twist components are shown.
We plot
$\sigma_T^{\tau}+\sigma_L^{\tau}$ for $\tau=2,4,6,8$, together with
the exact result (solid line).
The overall impression is that near $\xi=1$ all higher twist corrections 
are getting large, leading to the conclusion that the concept of higher twist
becomes meaningless.  If we naively extrapolate the higher twist
formulas  to the
region $\xi>1$ they diverge. 
Nevertheless, to the right of $\xi=1$ there is a region 
where twist 6 and 8 are small and can be neglected, 
whereas twist-4 accounts for the
deviation between the twist-2 approximation and the exact result.
Fig.~6 illustrates this in another way: the sum of twist-2 and twist-4
provides a rather accurate description of the exact formula. 
In this region 
(in Fig.3 between the two lines $\xi=0.2$ and $\xi=0.9$) twist-4 
corrections should improve the QCD description of deep inelastic
scattering.

In Fig.~7 we present the results of the  analysis performed only
for the longitudinal twist contributions $\sigma_L^{\tau}$ and $\tau=2,4,6,8$.
The striking result in comparison to the total cross section analysis
is a large and negative twist-4 contribution which accounts for a large
difference between the exact and leading twist result. 
This is shown  in
Fig.~8: again  the sum of twist-2 and twist-4 
provides an accurate description of the exact formula.
Notice that the difference $\sigma_L^{\tau=2}-\sigma_L^{exact}$ is large
already at $\xi=0.1$.  
Taking this result, we conclude that an analysis of the longitudinal
structure functions $F_L$ based entirely on the leading twist result 
is unreliable already for quite high values of $Q^2$ and not to small $x$. 
A similar effect, although not so pronounced, occurs for the transverse cross
section (large twist-4 corrections for the transverse case were also
discussed in \cite{MR}). The role of higher twist in the longitudinal structure
function was broadly studied in \cite{KS} and a qualitatively similar
results to ours were obtained there for  small $x$.

In summary, 
the main lesson to be learned from this study is the cancellation of higher 
twist in $F_2$: twist-4 is not small, neither in the transverse nor in the
longitudinal part, but it is hardly visible in the sum of both because
of the mutual cancellation.
Moreover,
 the twist-2 approximation to $F_2$ works even beyond $\xi=1$ where
the whole  concept of a twist expansion should make no sense.
As far as 
$F_L$ is concerned, the leading twist significantly exceeds the
exact result for $\xi>0.1$, and a large and negative twist-4
correction is necessary to obtain an agreement with the exact result.

\section{Discussion and Conclusions}

In this article we have carried out a simple numerical analysis of gluonic 
twist-4 corrections in the low-$Q^2$, small-$x$ HERA. We have reviewed
what an analysis of lowest-order QCD diagrams suggests. One of the most 
striking features are differences in sign between transverse and longitudinal
twist-4 corrections which may lead to a small twist-4 correction, even 
if the corrections to $F_T$ or $F_L$ are not small at all. Because of the
unknown initial conditions more input is needed. We then have used the
saturation model which provides an excellent description of both
$F_2$ and the DIS diffractive cross section at HERA, and we have analyzed 
its higher twist content. We found a one-to-one
correspondence between the twist expansion of this model and the QCD diagrams 
discussed before. The model can therefore be used to define initial conditions
of a QCD higher twist analysis.\\ \\
In our numerical analysis we have restricted ourselves to a careful study of
the saturation model. We found that, indeed, twist-4 corrections to $F_2$
remain small,  and  this smallness is due to an almost complete 
cancellation between large corrections to $F_T$ and $F_L$.  This implies 
that although twist four corrections are small in $F_2$, the use of the 
leading-twist DGLAP formalism for extracting structure functions becomes 
doubtful in the low-$Q^2$, small-$x$ HERA region. Within the saturation model
we have quantified the limit of applicability. 

This problem is even more 
acute for $F_L$ alone, where the large twist-4 correction  is
directly exposed, providing a crucial contribution which brings the leading
twist result close to the exact one. The DGLAP formalism might not be reliable
in this case even for higher values $x$ and $Q^2$, see Fig.~8.
\\ \\
Clearly, our numerical conclusions are based upon a specific model.
The phenomenological success of this model provides some reasons to believe 
that the conclusions are realistic. Moreover, on a qualitative level our 
conclusions are in agreement with the independent fit to the HERA data of 
~\cite{MRST}. Nevertheless, some uncertainty remains. We have outlined
that in the saturation model some contributions are not present which
one would expect to see when starting from QCD diagrams. If included they
may modify the subtle balance between transverse and longitudinal 
structure function. They may also shift the transition region
(the ``transition line'' of the saturation model, in reality, may turn out to 
be a rather narrow ``transition strip''). However, the coincidence with
the MRST fit makes us feel that the conclusions of our analysis are ``not far
from reality''.   

\newpage
\begin{figure}[t]
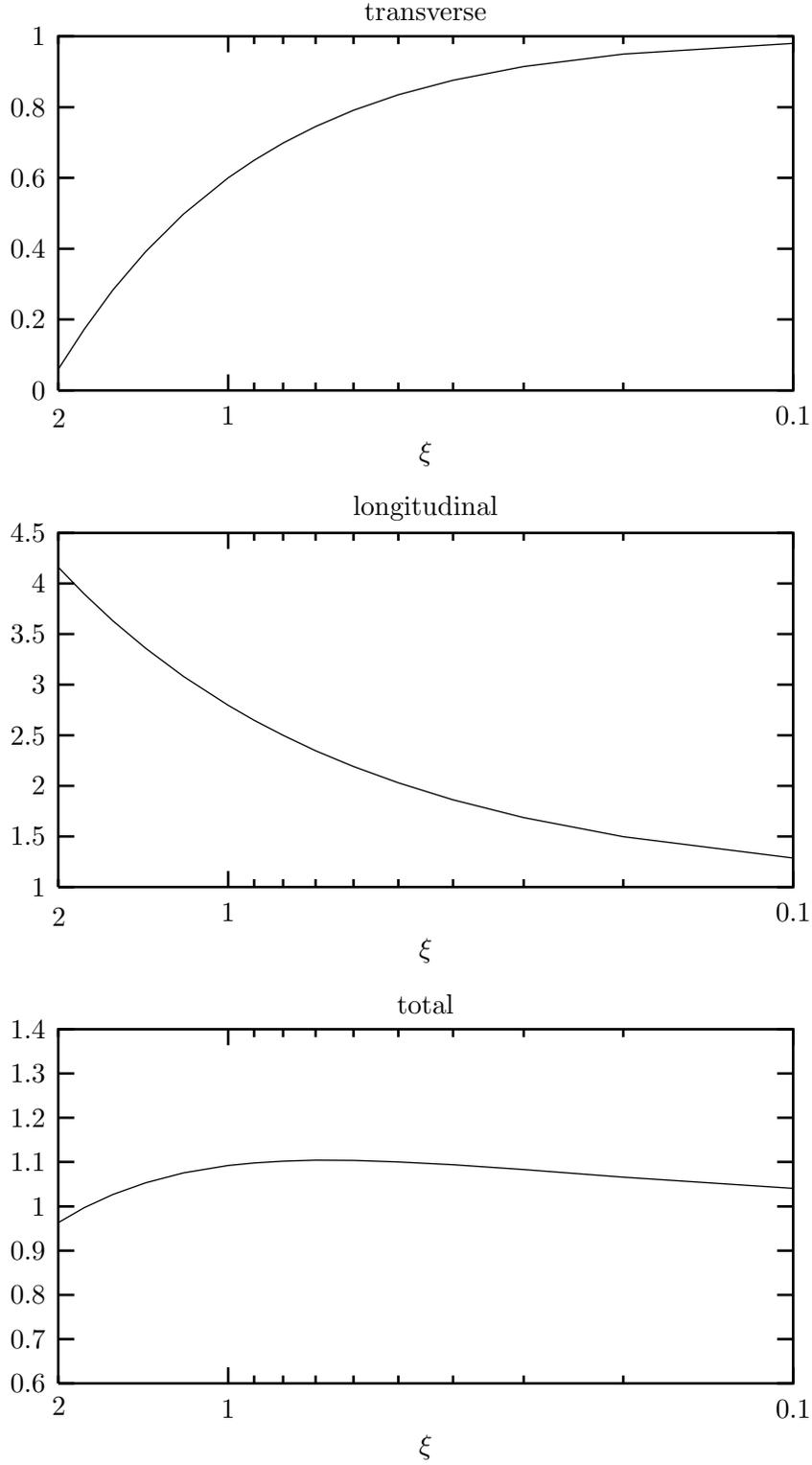

\label{figure4} 
  \begin{center} 
    \input{tw2t_twtot.pst} 
    \input{tw2l_twtot.pst} 
    \input{tw2_twtot.pst} 
       \caption{The ratio of the leading-twist and the 
                exact cross section in the saturation
                model : for the transverse, $\sigma_T^{\tau=2}/\sigma_T$,  
                longitudinal, $\sigma_L^{\tau=2}/\sigma_L$, and
                the total, $\sigma_{T+L}^{\tau=2}/\sigma_{T+L}$, 
                cross sections.}   
\end{center}  
\end{figure} 
\newpage
\begin{figure}[p] 
\label{figure5} 
  \begin{center} 
    \input{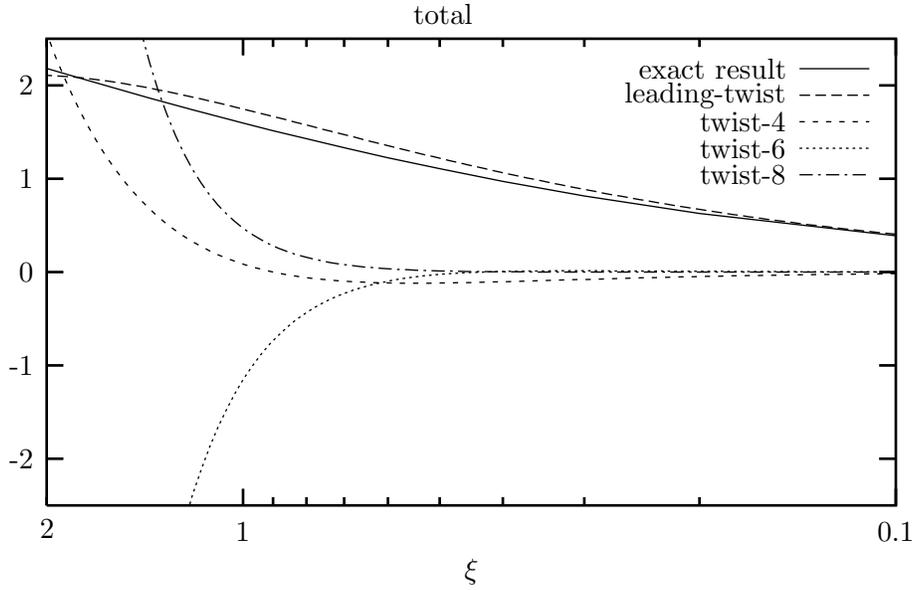} 
      \caption{The exact total cross section (solid line) and the higher
               twist contributions in the saturation model as a function of 
               the parameter $\xi$ defined in the text. The cross sections
               are rescaled by a common factor.} 
\end{center} 
\end{figure} 
\begin{figure}[p] 
\label{figure6}
  \begin{center}  
     \input{tw2+4.pst}  
       \caption{The same as in Fig.\ref{figure5} but with the leading twist
                and twist-4 added (dotted line).}  
\end{center} 
\end{figure} 
\newpage
\begin{figure}[p]
\label{plot7}
  \begin{center} 
    \input{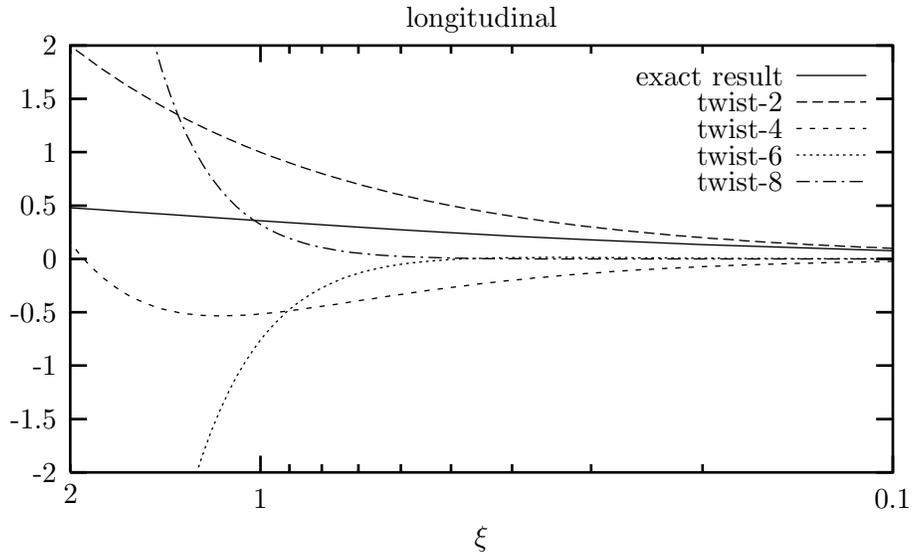} 
      \caption{The exact longitudinal cross section (solid line) and the higher
               twist longitudinal contributions in the saturation model 
               as a function of 
               the parameter $\xi$. The cross sections
               are rescaled by a common factor.} 
\end{center} 
\end{figure}
\begin{figure}[p]
\label{figure8} 
  \begin{center}  
     \input{tw2+4l.pst}  
       \caption{The same as in Fig.~7  but with the leading twist
                and twist-4 added (dotted line).} 
\end{center} 
\end{figure} 

\end{document}